# Automated Detecting and Repair of Cross-Site Scripting Vulnerabilities through Unit Testing

Mahmoud Mohammadi, Bei-Tseng Chu, Heather Richter Lipford

*Abstract*—The best practice to prevent Cross Site Scripting (XSS) attacks is to apply encoders to sanitize untrusted data. To balance security and functionality, encoders should be applied to match the web page context, such as HTML body, JavaScript, and style sheets. A common programming error is the use of a wrong type of encoder to sanitize untrusted data, leaving the application vulnerable. We present a security unit testing approach to detect XSS vulnerabilities caused by improper encoding of untrusted data. Unit tests for the XSS vulnerability are constructed out of each web page and then evaluated by a unit test execution framework. A grammar-based attack generator is devised to automatically generate test inputs. We also propose a vulnerability repair technique that can automatically fix detected vulnerabilities in many situations. Evaluation of this approach has been conducted on an open source medical record application with over 200 web pages written in JSP.

## I. Introduction

CROSS-SITE Scripting (XSS) is one of the most common security vulnerabilities in web applications. Cross Site Scripting attacks occur when an attacker successfully injects a malicious JavaScript payload into a web page to be executed by users requesting that page. Advised best practice to prevent XSS attacks is to encode untrusted program variables with dynamic content before their values are sent to the browser. The Acunetix Web Application Vulnerability Report [1] showed that nearly 38% and 33% of web sites were vulnerable to XSS attacks in 2015 and 2016 respectively. While one can prevent all XSS attacks by using the most strict encoder, that also takes away many useful web site functions. To balance security and functionality, developers must therefore choose the appropriate encoder depending on the context of the content, such as HTML or JavaScript. Research shows that as many as 28% of encoders are used incorrectly [2]. Static analysis [3] techniques are widely used to ensure a web application uses encoding functions to sanitize untrusted data. However, static analysis cannot verify whether the correct encoding function is applied.

Consider the fragment of a JSP program shown in Fig.1. Native Java code is enclosed in <% and %>. This example has two user-provided inputs: *pid* and *addr*. Variable *pid* is used as part of rendering an HTML anchor element on line 3, and *addr* is displayed in the HTML body on line 4. A maliciously supplied input for *addr* might be

```
<script> atk(); </script>
```

M. Mohammadi and B. Chu and H. Lipford are with the University of North Carolina at Charlotte

```
1  <% String pid=(String)request.getParameter("pid");%>
2  <% String addr=(String) request.getParameter("addr");%>
3  <a onclick="fn('<%=escapeHtml(pid)%>')" href="#" > mylink </a>
4  <p> <%=escapeHtml(addr) %>
```

Fig. 1: Motivation Example

If the encoding function, *escapeHtml()*, were not applied, the JavaScript function *atk()* on line 4 would be executed. The encoding function *escapeHtml()* replaces the < and > characters with < and > respectively and transforms the malicious input into the following string, preventing *atk()* from being interpreted as a JavaScript program by the browser:

<script> atk(); </script >

However, the same encoding function does not work for the case on line 3. A malicious input for *pid* might be the following:
'+ atk() + '

It will pass *escapeHtml()* unchanged. The rendered anchor element would be as follows:

```
<a onclick= "fn(''+ atk()+'')" href= "#" > mylink </a>
```

JavaScript function *atk()* will be executed as part of evaluating the input parameter expression of function *fn()* when the link is clicked. The correct JavaScript encoder would, in this case, replace the single quote character with escaped single quote \' to prevent this attack.

There are also cases where more than one encoding function must be used (e.g. an untrusted input used in both JavaScript and HTML contexts). The order of applying encoders is sometimes important as well. For example, Fig.2 shows a case in which the order of encoders is incorrect. The order is incorrect because the JavaScript encoder on line 3 is intended to prevent successful attacks by encoding single and double quote characters as they can be used to shape successful

```
1  <% user=
   request.getParameter("user");
2  var1 = escapeHtmlDecimal(user);
3  var2 = escapeJavaScript(var1)+
   "cnst"; %>
4  <a onclick="fn('<%= var2
   %>');">Details</a>
```

Fig. 2: Incorrect Order Of Encoders

attacks for the *onclick* attribute on line 4. However, in this case, the first encoder (*escapeHtmlDecimal*) replaces single quote characters with ' and this character combination will not be changed by the second encoder (*escapeJavaScript*). Thus, the encoded string by the first encoder ('); attack(); // ) can pass through the second encoder and be sent to the browser. Unfortunately, browsers will decode ('); attack(); // ) back to the original attack string ('); attack(); // ) leading to a successful attack. This vulnerability can be fixed by reversing the order of the encoders used.

Applying the correct encoding is thus context-sensitive, meaning the encoder must match the web element context where an untrusted variable occurs. In practice, a variable can occur in one of the following four contexts: HTML-body, JavaScript, CSS, and URL. Unfortunately, there is no systematic way to detect vulnerabilities due to mismatch of encoder and context [2]. Other researchers have looked at vulnerability prevention mechanisms using type inference to automatically detect the context of an untrusted variable so the correct encoding function can be automatically applied. To aid type inference, such efforts all rely on template languages with stronger type systems, such as Closure Templates [4] or HandleBars [5]. Such approaches have several limitations. First many web applications do not use such template languages. Second, type inference is not fully successful even with template languages. For example, a research team from Yahoo! found that they could identify the correct context in about 90.9% of applications written in HandleBars using type inference to detect the correct context. Other researchers have also shown that type inference is not always accurate for some program constructs written in Closure Templates [4].

Detecting XSS vulnerabilities through black box testing has also been researched, and there are several open source and commercial implementations [6]–[9]. In these approaches, a vulnerability is detected by inspecting web application outputs. If an injected attack payload is found in the output, the application is deemed vulnerable. However, this approach can lead to high false positives as an attack payload may not be executed by the browser. Black box testing could also have high false negative rates as well, due to inadequate test path coverage [6].

In this paper we present a unit testing based approach to automatically detect and repair XSS vulnerabilities due to incorrect encoding function usage. We have built a proof-of-concept implementation for web applications written in Java and JSP. This approach can be extended to other server-side web programming languages as well (e.g. PHP and ASP).

To detect XSS vulnerabilities we use an architecture composed of three components. First, to ensure XSS vulnerability test coverage, we construct multiple unit tests based on one given JSP file in the application. Second, we confirm each vulnerability by rendering attacked pages using a headless browser in a unit testing framework. Third, we have a structured way of generating attack strings as test inputs for unit test. Finally, to automatically fix detected vulnerabilities, we replace the vulnerable encoder(s) and re-evaluate using unit testing.

There are several contributions of this work. We minimize false positives by confirming vulnerabilities via execution in a real browser. False positives are a major obstacle for wide adoption of software security tools [6], [10]. Our testing approach can pinpoint exact locations of vulnerabilities, making it easy for remediation. We also minimize false negatives by ensuring path coverage for unit tests as well as systematically generating attack strings using BNF grammars based on modeling how browsers interpret JavaScript programs. Moreover, the proposed auto-fixing mechanism can fix many XSS vulnerabilities and use unit testing to either accept or reject the applied repairs.

In Section II we introduce an overall architecture of our approach and its components for XSS vulnerability detection and repair. We explain the unit test construction in Section III. Section IV explains an attack evaluation technique based on a unit testing framework followed by our grammar-based attack generation technique in section V. The vulnerability repair mechanism is introduced in section VI and section VII shows our evaluation results for vulnerability detection and repair. Section VIII reviews the related works followed by a conclusion and discussion of future research in section IX.

## II. ARCHITECTURE

The proposed approach combines static-dynamic vulnerability detection technique based on unit testing Static analysis is used to find the vulnerabilities have the advantage of complete source code coverage but suffer from high rate of false positive results due to ambiguities in determining whether a suspected vulnerability can actually be exploited. Dynamic analysis approaches can find the vulnerabilities with low rate of false positives due to using the real results of source code execution but suffer from the source code coverage issue leading to false negatives. to reduce both false positive and false negative. I use static analysis to maximizes code coverage to examine all execution paths. I use dynamic analysis to verify all attack inputs with a headless browser to minimize false positives. Our approach is designed to be integrated into unit testing frameworks such JUnit, so vulnerabilities can be mitigated early in the software development life-cycle.

Following our previous work [11], the overall architecture of our approach is shown in Fig.3. This figure lists the approach inputs as: the source code under the test and configurations including sensitive operations (security sinks) and untrusted sources. The output of this approach is a list of vulnerable points in the source code, many of them may be repaired by the auto-fixing component.

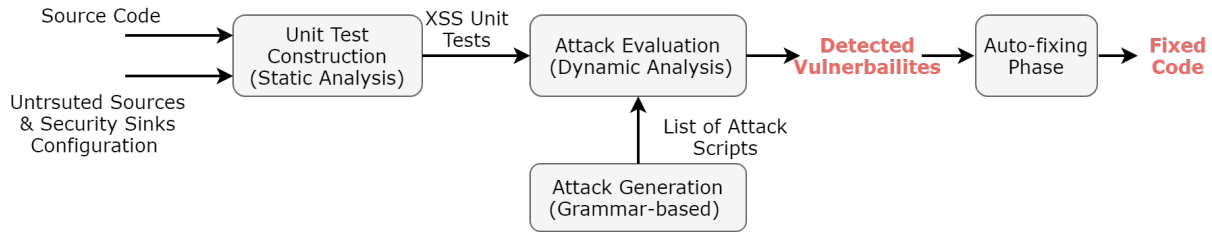

Fig. 3: Overall Architecture of XSS Unit Testing

Key components of XSS vulnerability detection include: XSS unit test extraction from application source, unit tests evaluation, and attack vector generation. I also proposed a automatic-fixing technique to repair many detected vulnerabilities using encoder placement.

The Unit Test Construction component analyzes the source code in order to automatically extract and generate unit tests for XSS detection in such a way to ensure test coverage. The Attack Evaluation component will use a repository of attack scripts, generated using the proposed Attack Generation component, to evaluate each unit test. Reported vulnerabilities may be examined by the auto-fixing component to find repairs.

Throughout the paper I will use the following terminology to explain our approach.

**Security Sinks** refer to program statements performing operations that could be subject to XSS attacks. Specifically, these server-side output generation commands sending values of variables to browser such as out.print() or <%= %> statements in JSP.

**Untrusted Sources** refer to statements retrieving information from sources that may contain malicious data. For example, *request.getParameter()* gets the data from the Internet. For our research, I assume untrusted sources are given as a set of API's returning untrusted values.

**Tainted Variables** These are variables that obtain their values (directly or indirectly) from the untrusted sources.

**Encoders** are functions that are used to generate safe version of their inputs using character encoding mechanisms such as *escapeHtml()*. Most developers use one of the widely used libraries of encoders such as ESAPI.

**Tainted Data Flow** is data flow of the tainted variable from an untrusted source to its destination in a security sink. I define each data flow as a tuple of untrusted source (U), security sink (S) and a list of encoders (E) between source and sink.

## III. UNIT TEST CONSTRUCTION

To ensure test path coverage, we construct a set of unit tests automatically based on each JSP file with the goal that if the original JSP file has an XSS vulnerability due to incorrect encoder usage, at least one of the constructed unit tests will be similarly vulnerable as well. We refer to the JSP file in the application as the original unit test and each unit test JSP file generated as the XSS unit test. The following are inputs for XSS unit test construction: (1) source code and (2) untrusted sources and (3) sinks.

We illustrate unit test generation using Fig.4 as the original code and Fig.5 as one of the constructed XSS unit tests. To focus our discussions, we assume the application encodes all untrusted variables using known encoding functions. Taint analysis can readily discover execution paths where an untrusted variable appears in a sink without encoding. Our vulnerability model is a situation where an encoder does not match the application's HTML document context.

```jsp
1  <% ord =request.getParameter("order");
2   ord = escapeHtml(ord);
3   if(editMode){ %>
4      <a onclick="edit('<%= ord %>')"
         href="#" > Edit Order </a>
5   <% } else { %>
6   <p> Order:<%= ord %> </p> <% } %>
```

Fig. 4: Original Source Code

Java variables and statements in a JSP file are referred to as host variables and statements. The term HTML document context refers to HTML and JavaScript code in the JSP file. To avoid false negatives, we test all sinks in all possible HTML document contexts. For convenience of performing program analysis, we replace all HTML elements with equivalent Java statements. This task is accomplished by using a JSP code analyzer that uses Java output generation commands such as out.write() to enclose HTML and JavaScript parts of the JSP files. For example, HTML elements on line 4 of Fig.4 are replaced by lines 5-7 in Fig.5.

Java branch statements could impact a sink's HTML document context as illustrated by Fig.4. Untrusted variable *ord* is in a JavaScript context in the "then" branch of a Java if-statement (line 4). In the "else" branch of the same Java if-statement, variable *ord* is in an HTML body context (line 6). To ensure we test sinks in all possible HTML contexts we test each of the Java if-statement branches in a separate unit test JSP file. Our intuition is that for most web pages, this will not lead to a large combinatorial explosion of test cases (see more in the evaluation section).

A control flow analysis is performed to generate control flow graphs for each JSP file. Multiple XSS unit tests will be created when the JSP file contains *if* or *switch/case* Java

```
1  String ord =
   request.getParameter("order");
2  ord= escapeHtml(ord);
3  //"then" branch of if statemenet
4  boolean e1= (editMode);
5  out.write("<a onclick=\"edit(' ");
6  out.write(ord);
7  out.write(" ')\" href=\"#\" > Edit
   Order</a>");
```

Fig. 5: Generated Unit Test

```
1  <% List<Profile> prf;
2  prf= searchProfile(customerID);
3  fName = escapeHtml(prf.Name); %>
4  <a onclick="profile('<%= fName %>
   ')" href="#" >
```

Fig. 6: Code with Untrusted Source from a Database call

```
1  //param is an input parameter
   containing an attack string at test
   time
2  param =
   request.getParameter("param");
3  prf= searchProfile(customerID);
4  // Injection point is in place of
   prf.Name in original code
5  fName = escapeHtml(param);
6  out.write("<a onclick=\"profile(\'");
7  // sink line in original code = 4
8  out.write( addLine(fName , "4") );
9  out.write(" ')\" href='#' >");
```

Fig. 7: Generated Unit Test with Injection Point

statements. For example, the source code of Fig.4 contains the following two possible execution paths:
- Line numbers 1,2,3,4 (then branch)
- Line numbers 1,2,3,6 (else branch)

Two XSS unit tests are generated for this example, each corresponds to one execution path containing no branching logic and each has a sink containing one untrusted variable. Execution paths without sinks or untrusted variables are discarded as they are not vulnerable to XSS attacks.

Fig.5 is a XSS unit test extracted from the "then-branch" of the Java if-statement in Fig.4. The untrusted variable in this unit test is *ord*, which appears in a sink statement (<%= %>). The sanitizing function is on line 2. To avoid any runtime exceptions or miss any statements affecting the HTML context we keep the conditional expression used in the if-statement in both branches, which is shown as on line 4 of Fig.5, by assigning the value of the conditional expression (editMode) to a Boolean variable e1.

While it is possible for branch statements written in JavaScript to change the HTML document context of a sink, we expect such cases to be rare. This is because sinks are written in Java. It is therefore natural for developers to use Java to express changes in HTML document context. We thus assume that JavaScript code does not change the HTML document context of sinks. We will assess this and other assumptions in the evaluation section.

We assume that each JSP web page is set up for unit testing. This means we have a runtime environment with web server, application server, and database server. Running XSS unit tests do not have additional requirements. The original JSP page is launched on a web server to set up the session. A proxy captures the session information. The captured session is sent as part of subsequent requests for XSS unit testing. The process described above is standard practice for unit testing web pages.

**Single Variable.** In the ideal case, the original JSP file contains one untrusted variable as is the case in Fig.4. For such a case, there are no false negatives because all possible HTML document contexts are captured by at least one XSS unit test. If the original code was vulnerable due to using the wrong encoding function, then at least one of the XSS unit tests would be vulnerable as well.

We define a false positive as a situation where the application's context and the applied encoding function are matched (safe) in the original source code, but the encoding function is detected as vulnerable (mismatched) by an XSS unit test. This is not possible for the ideal case because our XSS unit test construction process preserves the HTML document contexts of the original JSP file.

**Injection points for XSS unit tests.** We assume that untrusted sources are specified as a set of Java API's, such as user forms and database queries. Taint flow analysis is used to identify injection points in the program. Injection points are places where variables containing an attack string (as an input parameter for the unit test) are injected into a unit test. These variables are used as an argument of the first encoder function in its data-flow from untrusted source to security sinks. Since an XSS unit test contains no branching logic, detection of such injection points is straightforward. Fig.6 shows part of an original source code. Untrusted variable *fName* is used in a sink on line 4 after being sanitized using encoder on line 3. Variable *fName* gets value from variable *prf* as result of a database call, *searchProfile()*, a tainted source on line 2. In the corresponding unit test in Fig.7, variable containing the attack string *param* will be injected into the XSS unit test as the input parameter of the *escapeHtml()* encoder, as its first application in an statement, on line 3 of Fig.6.

We also instrument each XSS unit test so that it reports the line number in the source code if a vulnerability

```
1  <%= "User : " + escapeHtml(user) +
   "(" + escapeHtml(email) + ")" %>
2  <%= escapeHtml( "Patient:" +
   firstName + " " + lastName) %>
```

Fig. 8: Multiple Tainted Variables in one Sink

```
1  <div> <%= escapeHtml(request.getPara⌐
   meter("atk"))%>
   </div>
2  Attack String :  + alert(1)
3  <div>  + alert(1)   </div>
```

Fig. 9: False Positive in Attack Detection

```
1  Public void prepare() {
2  wt = new  WebTester();
3  sessionPreparation();
4  //other preparations such as proxy
5  }
6  public void run() {
7   for( String atk :atkVectors){
8   // Invoking the Unit Test
9   wt.gotoPage("unit1.jsp?param="+atk);
10  sleep(100);
11  verifyResponse(wt);
12  } }
```

Fig. 10: Test Driver with Test Preparation

```
1  var tags = document.all;
2  for (var i=0; i <tags.length;i++){
3   e= tags[i];
4   //finding events that has body
5   if (typeof e.onfocus == "function")
    {
6       event = e.onfocus;
7       e.onclick=event;
8       e.click();     }
9  // checking for other events of tag e
10 }
```

Fig. 11: Triggering all events having a handler

is found as shown on line 8 of Fig.7. We identify the line number of each sink statement in the original JSP file. Suppose the line number of a sink in the original JSP file is 4:

```
4: <%= tainted + "constant" %>
```

We add a function to each unit test to add the line number of the sink statement to the attack string:

```
out.write(addLine(tainted + "constant",4)))
```

Function *addLine()* is a server-side function which adds the line number of the sink statement as a parameter to the attack payload. This line number is calculated during the static analysis process generating XSS unit tests. In our evaluation described below, this line number will be used to identify the vulnerable statement line number to the developer and also used to guide the auto-fixing component to replace the incorrect encoders.

**Multiple Variables**: An XSS unit test may contain multiple untrusted variables. Fig.8 shows two examples. Best secure programing practices [12] suggest that if both variables are properly sanitized with respect to the expected HTML document context, their combination should be safe as well. We refer to this as the *independent encoding assumption*.

This assumption allows us to test one variable at a time by holding the rest of the untrusted variables constant. We will evaluate the assumption of encoding independence in the evaluation section.

## IV. ATTACK EVALUATION

The goal of attack evaluation is to assess whether an XSS unit test is vulnerable to any of the XSS attack strings. One widely used attack evaluation approach, exemplified by the popular black box testing tools such as NoScript [13], XS-SAuditor [14] and ZAP [9], is **String Matching Assessment** which aims to look up the attack payload in the response page. The rationale for this approach is that if an attack payload can bypass encoder functions intact, an attack could occur. Unfortunately this approach can lead to high false positives. A successful attack payload must be compatible with the context it is injected into. For instance, Fig.9 shows a situation in which an HTML body encoder is used to sanitize a user-entered parameter on line 1. Line 2 is an attack string from ZAP's attack repository. Line 3 shows a part of the output of the web page when this attack string is applied. Since the encoder does not alter the attack string, ZAP reports this page as vulnerable. This is a false positive because this attack cannot be executed in the HTML context.

Another approach known as **DOM Structure Assessment** is based on the observation that a successful attack can change the Document Object Model(DOM) structure of the response page. Assessing of this effect, known as lexical confinement, can be done by comparing the DOM structure of the response page using taint-aware policies written as taint tree patterns [15], [16]. Taint tree patterns are trees with regular expressions in their nodes describing different cases a successful attack can change the parse tree of an injected HTML node. This approach needs to define all possible taint tree patterns which can be very difficult especially for JavaScript codes, and leads to false negatives.

### A. Attack payload evaluation

Our approach is to execute unit tests by a headless browser such as JWebUnit. Vulnerabilities are only reported if successful execution of an attack payload by JWebUnit is detected [11]. For attack payload, we use a JavaScript function *attack(line)*, which takes as parameter the line number of the sink statement in the original source code being tested. The line number of each security sink is generated during the static analysis of the unit test construction as mentioned in the previous section. Function *attack()* changes the web page title by appending that line number. Web page title changes are monitored to detect successful attacks along with the location of the vulnerability in the source code.

### B. Test Driver

Fig.10 shows the XSS unit test driver. Lines 2 and 3 are for test preparation. Function *sessionPreparation()* sets up the execution environment by applying captured session information. The rest of the test driver invokes the XSS unit test by applying attack strings. After initializing an instance of WebTester (a subclass of JWebUnit) on line 2, each iteration of the loop on line 7 takes one attack vector (*atk*) and invokes the XSS unit test page ($unit1.jsp$) with the attack string as a parameter (line 9). Line 10 pauses to let the unit test page be rendered completely. Line 11 asserts whether the attack is successful by checking the title of the response page. If the attack is successful, the page title contains line number(s) of the vulnerable sinks, helping developers to fix vulnerabilities.

### C. Handling events

Attacks can be injected into vulnerable elements of HTML tags that are only executed upon triggering events, such as on key down. In order to find such vulnerabilities, we must trigger each event in the XSS unit test. There are 88 possible events in HTML5, some of them can only be triggered based on particular user interaction such as *onmouseover* or a run time condition such as *onerror*.

However, since all events share the same syntax, we can substitute events that cannot be easily simulated in a test environment with an event that can be easily triggered. We verified that in major browsers (Chrome, Safari, Firefox) event *onclick* can be associated with every HTML tag and it can be triggered using a JavaScript API. Fig.11 shows a JavaScript program we use to go through all tags in the DOM. For each tag, the program checks if the tag has an event with an event body (line 5). If a tag has a body, the program assigns the event body to an *onclick* event and triggers it automatically.

## V. ATTACK GENERATION

Because our test evaluation is based on execution of attack strings, we must make sure attack strings are syntactically correct. Furthermore, we want to include all possible types of attack scenarios. Related works in generating XSS attacks rely on either expert input [17], or on reported attacks [8], [18]. It is difficult to show that all possible attack scenarios are included using these approaches.

Our approach consists of two components. First we use context free grammar rules to model how JavaScript payloads are interpreted by a typical browser. Assuming they are accurate, then a successful attack must follow these grammar rules. Second, we devise an algorithm to derive attack strings systematically based on these grammar rules. Assuming the grammar rules accurately model the way the browser interprets JavaScript programs, and assuming that the attack derivation algorithm can generate at least one attack string for every type of attack, then our approach would cover all possible attack scenarios. It is possible that either we may have missed some grammar rules by which a browser interprets JavaScript programs, or the attack enumeration algorithm failed to consider a possible derivation path. Through peer review, we can improve both components in a way similar to any security algorithms are revised. The advantage of this approach is we rely expert know-how on the more manageable task of modeling browser behavior as opposed to the more open-ended task of enumerating possible attack scenarios.

### A. Browser Modeling

A typical web browser contains multiple interpreters: HTML, CSS, URI and JavaScript. The browser behavior can be modeled as one interpreter passing control to another upon parsing specific input tokens while rendering HTML documents. We refer to the event of interpreter switching as **context switching**. For example, the URI parser transfers the control to the JavaScript parser if it detects input *javascript:* as in the case:

```

```

A successful XSS attack is to induce the JavaScript interpreter to execute an attack payload. We use a set of context free grammar (CFG) rules to specify possible input strings that cause the browser to activate the JavaScript interpreter to execute an attack payload. Portners et. al. [19] observed that a successful XSS attack must either call a JavaScript function (e.g. an API), or make an assignment (e.g. change the DOM). According to JavaScript language syntax, wherever an assignment operation can be executed, a function call can also be made. Therefore, without loss of generality, we assume the attack payload (referred to as PAYLOAD in the following grammars) is a function *attack()* that changes the title of the web page.

Like Halfond et. al [20], we divide the CFG into these sections: URI, CSS, HTML, Event and JavaScript. In each section we specify possible transitions to cause a JavaScript interpreter to execute an attack payload. We will then integrate these sections of grammar rules to generate attack strings. For clarity, we will use the following convention in grammar definitions: upper case words for non-terminals, lower case words for terminals, symbols sq, dq, eq for single quote, double quote and equal sign characters respectively.

```
URIATRIB ::= URIHOST eq URIVAL
URIHOST ::= src | href | codebase | cite|action | back-
ground | data | classid | longdesc|profile |usemap | forma-
ction|icon | manifest | poster | srcset | archive
URIVAL ::= sq URI sq | dq URI dq | URI
URI ::= javascript: PAYLOAD
```

Fig. 12: URI Grammar

*1) URI context:* URI (Uniform Resource Identifier) strings identify locations of resources such images or script files. Based on RFC 3986, they have the following generic syntax:

scheme: [//[user:password@] host [:port]][/] path [?query] [#fragment]

Here, the scheme represents protocol type (such as ftp or http) used to access a resource, and the rest of the string expresses the authority and path information required to identify the resource. To cause the URI interpreter to switch to the JavaScript interpreter, the scheme must be equal to the keyword *javascript*, followed by JavaScript statements. Other possible schemes include http, ftp, and https. Since no JavaScript can be injected into schemes other than scheme javascript, we concentrate on describing URIs that contains the JavaScript scheme [21]. An URI can be properly interpreted by a browser only as a value of an expected attribute of a host context. We continue with the example of

``

where *src* is the source attribute of the HTML *img* tag and referred to as URIHOST. Fig.12 represents the grammar for URI. Rule URIATRIB specifies a URI attribute consisting of a URIHOST name and the URLVAL. Rule URIHOST lists all possible URI host contexts in an HTML document. Again, for the purpose of generating attack strings, we only consider a URI of the JavaScript scheme. PAYLOAD is a special nonterminal representing a JavaScript attack payload. It signals to the attack generator that a context switch to JavaScript is possible at this point.

*2) CSS Context:* Cascading Style Sheets (CSS) specifications can be either contained in a CSS file or placed directly in HTML elements, e.g. tag definitions (using the *style* attribute or style blocks). A context switch from the CSS interpreter to the JavaScript interpreter is possible only when a URI is a property of a CSS-style element, specified by function *url()*. The argument to the *url()* function must follow the definition of URI in Fig.12. Fig.13 lists rules for URI to be included as part of a CSS-style element.

*3) Attribute Event Context:* HTML events, such as onfocus and onload, can cause context switches to JavaScript. Grammar rules in Fig.14 define an HTML event attribute composed of an event name EVENTNAME and value EVENTVAL. Although types of possible events vary with HTML tags, we found that the *onclick* event can be triggered in all HTML tags. As mentioned in attack evaluation section,

```
STYLEATRIB ::= style eq STYLEVAL
STYLEVAL ::= (sq STYLE sq) | (dq STYLE dq) |
(STYLE)
STYLE ::= CSSPROP*
CSSPROP ::= PROPNAME : PROPVAL;
PROPNAME ::= background-image | list- style-image|
content | cursor | cue-after | cue-before
PROPVAL ::= url(URI)
```

Fig. 13: CSS Grammar

```
EVENTATRIB ::= EVENTNAME eq EVENTVAL
EVENTNAME ::= onclick
EVENTVAL ::= sq PAYLOAD sq | dq PAYLOAD dq |
PAYLOAD
```

Fig. 14: Event Attributes Grammar

```
HTML ::= ELEM*
ELEM ::= IMG | STYLE | SCRIPT | SPECIAL
IMG ::= 
ATRIBLIST ::= ATTRIBUTE*
ATTRIBUTE ::= URIATRIB | STYLEATRIB | EVEN-
TATRIB
STYLE ::= <style> CSSPROP* </style>
SCRIPT ::= <script> PAYLOAD </script>
SPECIAL ::= ( </textarea> | </title> )
```

Fig. 15: Integration Grammar

we change all events in the source code to the *onclick* event for attack evaluation. Rule EVENTVAL defines the value of the event which is a JavaScript statement to be executed upon the specified event.

*4) HTML:* Having modeled context switches in URI, CSS, and Event, we integrate them in a single grammar to model JavaScript execution in HTML as shown in Fig.15. A XSS attack script can be injected either in a tag's attribute or tag's body. Rule HTML in Fig.15 defines tags as a set of elements represented by the ELEM rule to cover these cases.

Since all HTML tags attributes share identical syntax, we use rule IMG to define tag *img* as a representative to model all possible context switching patterns via tag attributes. The browser can switch to the JavaScript interpreter only in the following tag attribute types: URI, CSS, and EVENT. Grammar rules for these elements have been discussed above.

In the case of injection into tag bodies, JavaScript must be enclosed by the <script> </script> tags, as specified by the SCRIP rule. However, there are a few exceptions. First, inside <style> tag body, JavaScript can only be included as part of some CSS properties, as specified by STYLE rule. Second, no JavaScript are allowed in bodies of <textarea> and <title> tags. To inject JavaScript into bodies of these tags, these tags must first be closed as specified by rule SPECIAL.

```
ADDITIVEXP ::= PRIMARYEXP ADDITIVEPART
ADDITIVEPART::= (+ PRIMARYEXP)*
PRIMARYEXP ::= PAYLOAD | LITERAL
LITERAL = dq 1 dq | sq 1 sq | 1
```

Fig. 16: JavaScript Additive Expressions Grammar

*5) JavaScript:* JavaScript code can be placed either directly in HTML elements (e.g. through tag events such as *onclick*) or in <script> blocks. Attackers can inject a malicious payload into a block of vulnerable JavaScript code. A successful attack must manipulate the JavaScript interpreter into executing the payload, *attack()*.

Injection points in JavaScript are (Java) host variables. While host variables could be used in any JavaScript construct, such as part of a variable or function name ( e.g., `var vname<%= hostVar %> = 'value';`) such cases make little sense. Host variables are primarily used to pass server-side values to JavaScript code. Thus we only consider scenarios where attack scripts are injected as part of a string or a numeric literal in expressions as illustrated by the following examples in a JavaScript block.

```
1  var x = " const <%= hostVar %> " ;
2  var x = 19<%= hostVar %>;
3  var x = 20 * <%= hostVar %>;
4  func("const" + <%= hostVar %> ,
    param2);
5  if ( <%= hostVar %> == 2017) {...}
```

The goal of the each attack script is to turn the host variable into an expression so a function call can be made. A successful attack can be any syntactically correct JavaScript expression. Without loss of generality, we generate attack expressions using only the plus(+) operator as it can be used on both string and numeric data. The resulting expression is referred to as an additive expression. Its grammar is shown in Fig.16. The first two lines in Fig.16 define JavaScript additive expressions as expressions composed of multiple string/numeric literals or expressions concatenated to each other using the plus(+) operator in JavaScript. PAYLOAD non-terminal is a placeholder for attack payloads.

### B. Attack String Generation

The goal of the attack string generation is to generate all possible types of attacks using the grammar rules described in the previous section. We describe the generation process in this section.

*1) Sentence Derivation:* We generate XSS attack strings based on any of the grammars described above by constructing a leftmost derivation tree [22] from the start symbol of each grammar. The following are derivation steps for a sentence based on the HTML *img* tag grammar.

ELEM ::= IMG
 ::= 
 ::= 
 ::= 
 ::= 

*2) Generating Attack Strings:* Attacks can be injected in any part of an HTML element, a CSS block, or JavaScript expression. Consider the following example where a host variable, *hostVar*, is a function parameter on the right hand side of the assignment statement for variable *fName*.

```
var fName =func("Dr. <%= hostVar %> ");
```

The attack script must take into consideration existing characters both to the left and to the right of the injection point (point in which the *hostVar* is placed), referred to as left context and right context respectively as described in our previous work [11].

To fit the attack into the left context, one may close the string parameter with character " followed by a context switch using a new additive expression. The resulting attack string would be: **" + attack() + "** and the successful injection is shown as follows:

```
var fName =func("Dr." + attack() + "");
```

We first derive a sentence based on the start symbol of the grammar. Each sentence will lead to successful execution of a JavaScript attack. To systematically generate attacks for all possible existing left and right contexts, we must produce all possible partial sentences. The following is a possible derivation for an additive expression in JavaScript leading to a complete sentence:

```
ADDITIVEXP ::= PRIMARYTEXP ADDITIVEPART
. . . ::= LITERAL ADDITIVEPART
. . . ::= "1" ADDITIVEPART
. . . ::= "1" (+ PRIMARYTEXP)*
. . . ::= "1" + PRIMARYTEXP + PRIMARYTEXP
. . . ::= "1" + PAYLOAD + "1"
```

For each complete sentence derived from the grammar, we generate multiple versions of partial sentence as potential attack strings. Each version will be shaped by removing one token from the either the beginning or from the end of the previous version starting from the initial sentence. These versions represent different possible ways an attack can be successfully interpreted by the browser taking advantage of the injection point's left and right contexts.

This removing process will continue until the first PAYLOAD symbol is reached. For example, given the additive expression derived earlier, the following versions of attack strings can be generated.

1) "1" + PAYLOAD + "1"
2) 1" + PAYLOAD + "1"
3) " + PAYLOAD + "1"
4) + PAYLOAD + "1"

5) PAYLOAD + "1"

To consider existing contexts to the right of the injection point, we systematically generate multiple versions of any partial attack string by removing one token from the end of the previous one until the PAYLOAD symbol is reached. The following four versions of attack strings are derived based on attack string 3 from previous list:

6) "+PAYLOAD + "1
7) **"+PAYLOAD + "**
8) "+PAYLOAD +
9) "+PAYLOAD

Attack string in item 7 can be successfully injected into host variable *hostVar*.

*3) Closures:* Closure operators (*, +) in our grammar rules may result in an infinite number of derivations. The following example shows a derivation by applying the closure operator up to two times on the ELEM rule. A total of six derivations are possible for the ELEM non-terminal:

HTML ::= ELEM*
ELEM ::= (IMG | SCRIPT)*
::= IMG
::= IMG IMG
::= IMG SCRIPT
::= SCRIPT
::= SCRIPT SCRIPT
::= SCRIPT IMG

We observed that attack strings containing more than one attack payload are redundant. This is because a successful attack only needs to execute one payload. If an attack pattern is not successful, repeating the same pattern multiple times will not help it succeed. We empirically determine the number of times closure operators need to be applied. We compute leftmost derivations by applying different upper bounds on closure operators until no more attack strings with one payload can be added. For example, for the grammar rules presented here, applying each closure operator 3 times does not generate new attack strings with one payload over applying each closure operator 2 times. Note that current grammars do not contain recursive rules.

In summary we generate a set of attack strings by deriving sentences from the start symbol of each grammar (URI, CSS, HTML, EVENT, JavaScript). Each closure operation is applied up to two times. For each attack string in the initial set, we generate additional versions of the attack string by dropping tokens from both the left and right as described above. Only attack strings with a single context-switch are included in the final set of attack strings for unit testing.

## VI. AUTOMATIC REPAIR

After detecting vulnerabilities, the repair phase aims to automatically fix discovered vulnerabilities by replacing the

```
<% String user =
request.getParameter("username");
user = escapeHtml(user); %>
User Name: <div ><%= user %> </div>
<img src="plus.gif"
onclick="details('<%= user %>')" >
```

Fig. 17: Original Source Code

```
1) <% String param = JavaScriptEnco
der(request.getParameter("param"));%>
<script> Func( 9<%= param %>);
</script>
// Attack Script : + attack();
2) <script> window.setInterval('<%=
param %>');</script>
// Successful Attack Script:
attack();
```

Fig. 18: Proper Encoders but Vulnerable

incorrect encoders with proper ones. Consider the sample code snippet of Fig.17, security unit testing will reveal an XSS vulnerability on line 4 as the HTML encoder used on line 2 will not prevent XSS attacks in the JavaScript context. The following is an example of a successful attack against this vulnerability :

'); attack(); //

We observed that there are four choices of contexts of base encoders: HTML, JavaScript, URL and CSS. One must also consider combinations of multiple encoders. As we discussed in the introduction, a wrong order of encoders can also lead to vulnerabilities [23]. The OWASP secure programming guideline [24], a highly regarded source for secure programming, suggests the following six possible encoders and their combinations: HTMLEncoder, JavaScriptEncoder , CSSEncoder, URLEncoder , JavaScript(HTML()) and JavaScript(URL()) as adequate for preventing the vast majority of XSS vulnerabilities. We refer to this list of six choices as *candidate encoders*.

We explore the possibility of automatically fixing XSS vulnerabilities by trying each of the possible candidate encoders to replace vulnerable encoders and use the attack evaluation mechanism described in section IV to verify if the replacement produces a program not susceptible to XSS attacks. This repair strategy is computationally feasible for most program structures due to the limited number of candidate encoders (6 encoders) and the short time required to verify each encoder replacement. However, there are a few caveats to the auto fixing approach.

First, we may be able to fix a vulnerability but unintentionally lead to unexpected behavior. For example, a *JavaScript(Html())* encoding sequence may be applied to encode a variable (e.g., *companyName*) inside a JavaScript block. If the value of *companyName* happens to be safe value of Johnson & Johnson, what would be displayed to the end user could be Johnson ∓ Johnson. We refer to this problem as *over-encoding*. It is very difficult to avoid over encoding as there is no precise definition. We can minimize the likelihood of over-encoding by considering encoders in the candidate encoder list. Furthermore, we choose repairs with single encoder over repairs that involve double-encoder.

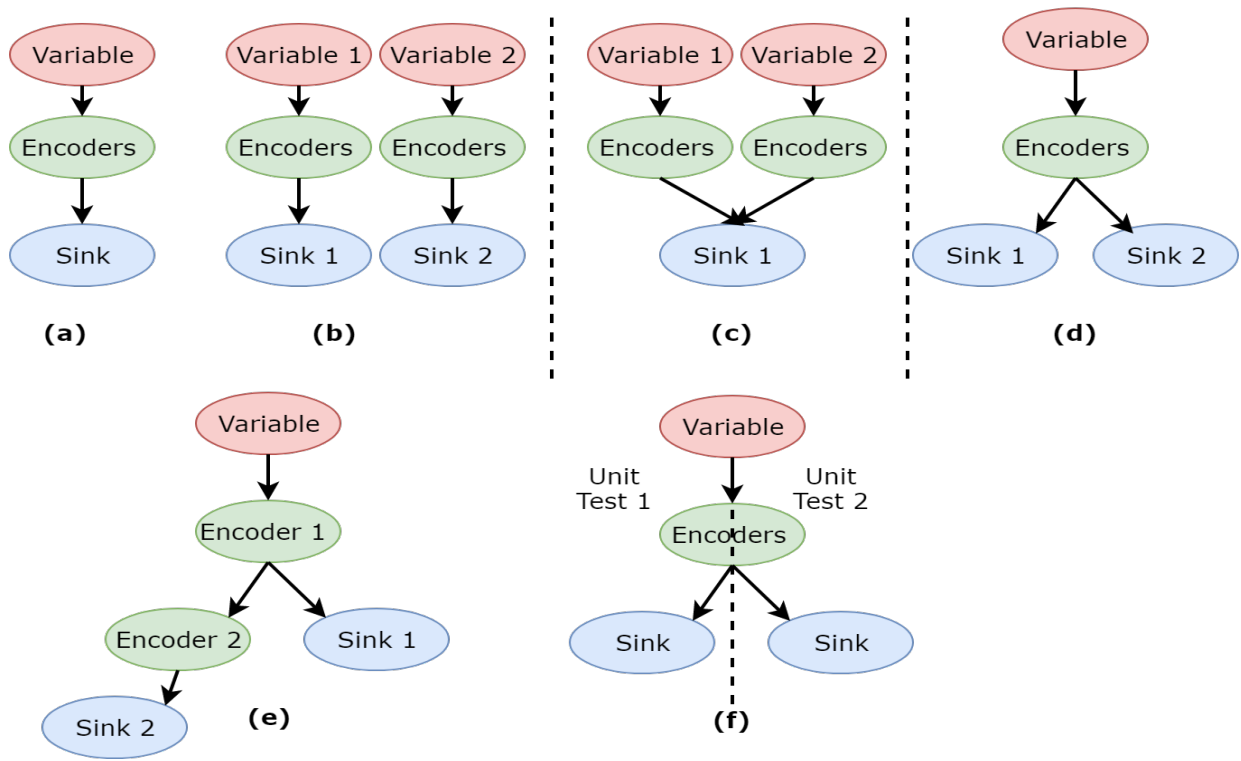

Fig. 19: Different Cases of the Automatic Repair: (a) Single Variable, Single/Multiple Encoders. (b) Multiple Variables (c) Multiple Variables -Single Sink (d,e) Multiple Sinks - Shared Encoder. (f) Multiple Unit Tests

Second, popular encoding libraries such as Apache, Spring framework, and ESAPI, differ in implementation details that can cause vulnerabilities, as illustrated on lines 1 and 2 of Fig.18. On line 1, a JavaScript encoder should be used to sanitize variable *param*. Both Apache and Spring libraries did not prevent the attack string listed on line 3, but the JavaScript encoder from ESAPI library is safe. The reason is that ESAPI encodes character plus(+) while the the other two leave it unchanged.

Third, it is possible that no fix can be found by using one of the candidate encoders. In such cases, we will defer the fix to developers. For example, we cannot fix unsafe programming practices outlined by OWASP, as illustrated on lines 2 and 4 of Fig.18. JavaScript API *setInterval()* is inherently unsafe because it may take *attack()* directly as an argument. No encoders can fix this vulnerability. We do not consider repairs that require structural changes to the program, like adding a new variable, add or deleting statements.

Finally, we only vulnerable code where up to two encoders are used in a sequence, which should cover the vast majority of cases [25] Vulnerabilities with more than two encoders in a sequence are referred to the developer as auto-fixing is likely to lead to over encoding.

The core task for repair is to replace the vulnerable encoder and perform XSS unit testing to either accept or reject the fix. Fig.19 shows all possible scenarios for encoder replacement. We examine each scenario in detail.

**Single Variable:** In this case a tainted data-flow only

```
1  <% user=
     request.getParameter("user");
2  user= escapeHtml(user); %>
3  <a onclick="fn('<%= user
     %>');">Details</a>
```

Fig. 20: Single Variable, Single Encoder

```
1  <% user=
     request.getParameter("user");
2  user = escapeHtmlDecimal(user);
3  user2 = escapeJavaScript(user)+
     "constant"; %>
4  <a onclick="fn('<%= user2 %>');">
     More Details</a>
```

Fig. 21: Single Variable, Multiple Encoder

```
1  <% user =
   request.getParameter("user");
2  user = escapeHtml(user); %>
3  <a onclick=" details('<%= user %>
   ');" > Details </a>
4  <%
   email=request.getParameter("email");
5  email = escapeHtml(email); %>
6  <a onclick=" fn(' <%= email %> '); "
   > Send </a>
```

Fig. 22: Multiple independent tainted variables

```
1  <% user =
   request.getParameter("user");
2  user = escapeHtml(user);
3  email =
   request.getParameter("email");
4  email = escapeHtml(email);
5  fullusr = user+"(" + email +")"; %>
6  <a onclick=" fn('<%= fullusr %>'); "
   > Details </a>
```

Fig. 23: Multiple Variable - Single Sink

contains one tainted variable and the tainted variable contains up to two encoders in its path from the untrusted origin to the sink as shown in Fig.19(a). Fig.20 shows a sample code snippet for this case in which the tainted variable *user* has been used in a sink on line 3 after being sanitized using encoder *escapeHtml()* on line 2.

Repairing this vulnerability entails replacing the vulnerable encoder (*escapeHtml*) on line 2 with another one from the list of *candidate encoders*. After modifying the code with each of the candidate encoders the modified code should be tested again. In this example, encoder (*escapeJavaScript(escapeHtml)*) would fix the vulnerability. The computational complexity of this case is the time required to test all the candidate encoders.

Fig.21 shows an example where an untrusted variable is sanitized by two encoders on lines 2 and 3 before sending its value to the browser via the sink statement of line 4. This code is vulnerable because of order of encoders. Encoder *escapeHtmlDecimal()* on line 2 replaces the single quote character with its decimal equivalent which can bypass the JavaScript encoder and be sent to the browser unchanged. Once decimal encoded characters are parsed by the browser they will be decoded back to original single code leading to a successful attack as mentioned in introduction section.

To find a solution for this two-encoder case, we test the following combinations from the candidate list where e1 and e2 refer to encoders on lines 2 and 3 respectively.

- { e1: escapeJavaScript , e2: escapeHtml }
- { e1: escapeJavaScript , e2: escapeURL }

This single-variable case also covers situations in which two encoders are nested in one statement such as the code below:

```
user = escapeJavaScript( escapeHtml(user));
```

The single-variable case can be generalized to situations where a unit test contains multiple independent data-flows as shown in Fig.19(b). Fig.22 shows a unit test that contains two tainted variables (*user* and *email*), they are used in independent different sinks (line 3 , 6) with separate encoders on lines 2 and 5. In this example, both encoders are incorrect. Our approach can automatically fix these vulnerabilities by replacing both encoders as *escapeJavaScript()*. Because the vulnerable sinks have independent data-flows they can be evaluated at the same time. The computational complexity for cases in Fig.19(a) and (b) are the same.

**Multiple Variables - Single Sink:** In this scenario one security sink is the end point of multiple untrusted variables with separate encoders in their data-flows as shown in Fig.19(c). A vulnerability is reported if at least one of the encoders is incorrect. Fig.23 shows such a case in which two tainted variables *user* and *email* are concatenated to shape the third variable *fulluser* to be used in the sink on line 6 after *user* is sanitized on line 2 (refereed as e1) and *email* on line 4 (refereed to as e2).

We observe that in most cases, variables in a given sink appear in the same web application context. This implies all variables should use same encoders. In the example of Fig.23, because variables *user* and *email* appear in the same context (i.e. JavaScript argument in an event), considering the following replacements are sufficient.

1) { e1: escapeJavaScript , e2: escapeJavaScript }
2) { e1: escapeHtml, e2: escapeHtml }
3) { e1: escapeCSS , e2: escapeCSS }
4) { e1: escapeURI , e2: escapeURI }
5) { e1: escapeJavaScript (escapeHtml()) , e2: escapeJavaScript (escapeHtml()) }
6) { e1: escapeJavaScript (escapeURI()) , e2: escapeJavaScript (escapeURI()) }

However, one could imagine rare cases where multiple variables in one sink may be rendered in two or more contexts. For such cases, we must consider testing replacements where e1 and e2 are different, or 6*6=36 encoder combinations. This would be computationally expensive if many variables are involved. We believe such cases are rare. So our proposal is to only test the same encoder sequence for all variables at the same time. If a repair cannot be found, this may indicate multiple contexts are involved for the same sink. We defer repair for such vulnerabilities to developers.

**Multiple Sinks - Shared Encoder:** These are cases where different sinks share the same set of encoders as shown in Fig.19(d) and (e). A vulnerability appears when a sink's context does not match the shared encoder. Fig.24 shows

```
1  <% user =
   request.getParameter("user");
2  user =escapeHtml(user); %>
3  <a onclick= " Add( ' <%= user %> ' )
   " > Add </a>
4  <a onclick= " Edit( ' <%= user %> '
   ) " > Edit </a>
```

Fig. 24: Multiple Sinks-Shared Encoder

```
1  <% user =
   request.getParameter("user");
2  user =escapeJavaScript(user); %>
3  <p> <%= user %> </p>
4  user =escapeHtml(user);
5  <a onclick= " Add( ' <%= user %> ' )
   " > Edit </a>
```

Fig. 25: Shared Encoder: Different Contexts with solution

```
1  <% user =
   request.getParameter("user");
2  user =escapeJavaScript(user); %>
3  <a onclick= " Add( ' <%= user %> ' )
   " > Add </a>
4  user =escapeHtml(user);
5  <p> <%= user %>   </p>
```

Fig. 26: Shared Encoder: Different Contexts and no solution

```
1  <% user =
   request.getParameter("user");
2  user =escapeHtml(user)
3  if(editMode){ %>
4  <a onclick="fn('<%= user %>')" >
   Edit User </a>
5  <% } else { %>
6  <div> User Name : <%= user %> </div>
   <% } %>
```

Fig. 27: Share encoder in multiple unit tests

such a case in which the sinks on lines 3 and 4 use the same encoder of line 2. To fix this vulnerability, the encoder on line 2 need to be replaced by

```
user=escapeJavaScript(escapeHtml(user))
```

Moreover, a developer may add an extra encoder before one of the sinks as line 4 of Fig.25. The more general pattern for this case of multiple sinks sharing common encoders is shown in Fig.19(e). This code is vulnerable because the encoder on line 2 does not prevent attacks to line 3. Using the list of candidate encoders, the repair found for the encoders on lines 2 and 4 would be:

Line 2: `user = escapeHtml(user)`
Line 4: `user = escapeJavaScript(user)`

However, there are situations where no repair can be made for this pattern of code. Consider the example in Fig.26 where encoded variable *user* is used in two different contexts: JavaScript on lines 3 and HTML on line 5. The code is vulnerable and a repair cannot be found for encoders on lines 2 and 4. The reason is that none of the OWASP two-encoder combinations ( { Line 2: HTML , line 4: JavaScript} or { Line 2: URL , Line 4: JavaScript} will lead to safe code.

To repair this vulnerability, a new variable will have to be created, changing the structure of the program. Our current approach does not consider such moves. Future research is needed to thoroughly explore this strategy.

**Multiple XSS Unit Tests:** So far we have considered possible scenarios to repair a vulnerability within a single XSS unit test through encoder replacement. We consider next situations where vulnerabilities are discovered in two different XSS unit tests derived from the same JSP page as illustrated in Fig.19(f). As long as fixes for each XSS unit recommend same replacements , the final fix for the JSP page can be easily constructed. An example of such a case is illustrated in Fig.27. In this case, each XSS unit test is based on a different branch of *if/else* statements. Lines 1,2,3,4 are in one XSS unit test and lines 1,2,3,6 are in another one. Line 2 contains the shared encoder between the two unit tests. Similar to the shared encoder in Fig.25, the correct encoder on line 2 should satisfy two contexts, as in:

`cmp = escapeJavaScript( escapeHtml(cmp))`

However, it is possible that there is a conflict in repairs for each XSS unit test. In such a situation structural changes to the code is required by a developer to fix this vulnerability.

## VII. EVALUATIONS

Our evaluations use iTrust, an open source medical records application with 112,000 lines of Java/JSP code [26]. Project iTrust has 235 JSP files and we use all of them for this evaluation. We seek to evaluate the following research questions.

**(1)** Are the assumptions made in our approach valid in iTrust?
**(2)** How effective is the described approach at detecting XSS vulnerabilities?
**(3)** How does XSS Unit-Testing compare with existing tools in detecting vulnerabilities?

**(4)** What is the computational performance of the described approach at detecting XSS vulnerabilities?
**(5)** How effective is the describe approach at auto-fixing detected vulnerabilities?

### A. Assumption Verification

We assume that all web pages can be executed in a unit test environment without runtime errors. This implies that all resources required to run these web pages, such as application servers, database servers and external libraries are available for both vulnerability detection and repair phases. These requirements are met with the iTrust project. Each of the 235 JSP pages can be executed successfully as unit tests. We use the Apache TomCat as application server and mySQL server as the database. iTrust uses Apache StringEscapeUtils libraries to encode the outputs and traditional JSP tags to generate outputs.

We assume that untrusted variables are independent of each other. This means that if a unit test contains more than one variable that may contain malicious input, we can find all XSS vulnerabilities by testing each variable independently. Out of 2268 sinks in iTrust, 27 contain multi-variables. In all these cases our encoding independence assumption is true.

We also assume that the JavaScript codes do not change web context of sinks and server-side variables are only used as values in JavaScript programs. We found these assumptions are true in all cases in where server-side values are passed to JavaScript blocks.

### B. Vulnerability Detection

We compared our XSS unit testing approach with security black box testing using a popular open source security testing tool ZAP [9]. Table I summarizes our evaluation results.

TABLE I: Summary of vulnerability findings

|  | Detected Vuln. | True Positives | False Positives |
|---|---|---|---|
| ZAP | 119 | 10 | 109 |
| XSS Unit Tesing | 24 | 24 | 0 |

We found 24 zero-day vulnerabilities due to misuse of encoders. The following code snippet provides an example from iTrust where HTML encoding is used in a JavaScript context.

```
<a onclick="func('<%= escapeHtml(input)
%>')" > Link</a>
```

ZAP has a very high false positive rate: 91%. No false positives were reported by our approach. The reason for ZAP's high false positive rate is because it does not confirm findings through execution. Instead it uses string matches to find attack scripts in output pages, as illustrated in section IV.

Our approach found 14 vulnerabilities ZAP did not find. All these cases are due to the lack of test coverage by ZAP. ZAP does not test all execution paths. In our approach, a separate XSS unit test is created for each possible execution branch in a JSP file. In addition, some vulnerabilities are triggered by events, such as failure to load an image. Our test evaluation approach handles such situations.

### C. Attack Generation

We compared our grammar based attack generation with two well regarded open source XSS attack repositories: ZAP repository and the HTML5Sec web site [17]. The HTML5Sec attack repository found fewer vulnerabilities than the ZAP repository. However, we found several vulnerabilities that cannot be detected by ZAP or HTML5Sec repositories. One example is shown below.

```
<div style="height: <%= escapeHtml(inpu
t) %>px; "> </div>
```

The following attack string generated by our approach can detect this vulnerability.

```
;background-image:url('javascript:atk()
');
```

Attack repositories in ZAP and HTML5Sec rely on contributions from pen-testing experts. Our approach systematically derives attack strings based on a set of grammar rules modeling the behavior of browsers interpreting JavaScript programs.

### D. Computational performance

We looked at the performance of XSS unit testing using experiments performed on a desktop Mac with a 2.7 GHz Intel core i5 with 8GB RAM. Our attack generator produced 223 attack strings, which were applied to each unit test. For iTrust, it takes 17 seconds on average to evaluate a XSS unit test. A JSP file may contain multiple branches of execution paths but only those containing sinks with tainted variables will be tested. Our evaluation of 235 JSP pages in iTrust shows that on average a JSP file leads to 29 XSS unit tests. On average, if a JSP page contains no vulnerabilities, our approach will take 493 seconds or 8.2 min to complete all the unit tests. Generation of XSS unit tests is much faster than running all the tests. Because each JSP file can be tested independently, this approach lends well for parallel processing. Overall, we believe the approach we described in this paper may scale well for large applications.

### E. Auto-Repair

We applied the described auto-fixing mechanism to all 24 vulnerabilities found in iTrust. Our approach is able to automatically fix all of these vulnerabilities. Fig.28 shows an example of a vulnerability on line 1 and its repaired version on line 2. Line 1 shows a vulnerability due to incorrect use of a HTML encoder (*escapeHtml()*) for JavaScript context (*onclick* event attribute) for untrusted variable *tempName*. This vulnerability can be exploited using an attack script like **'+ attack() + '**.

All iTrust vulnerabilities are of the pattern (a) and (b) in Fig.19. The time required to evaluate each candidate encoder is the same as the time required to evaluate the unit tests for the vulnerability detection phase. On average it takes testing for two candidate encoders before a fix is found.

```
1  <a onclick= "fn('<%=
   escapeHtml(user) %>')" > ... </a>
2  <a onclick= "fn('<%=
   escapeJavascript( escapeHtml(user ))
   %>')" >...</a>
```

Fig. 28: Vulnerable Code and Repaired Version

## VIII. RELATED WORK

**Vulnerability Detection:** Researchers have investigated a variety of prevention and detection techniques to mitigate XSS vulnerabilities. Preventive approaches include secure programming guides to inform developers how to use encoding functions correctly. Well known guidelines include the OWASP XSS cheat sheet [24] and best practices by Graff and Wyk [27]. Attempts have been made to automatically sanitize untrusted inputs using template languages. As we discussed in the I, approaches for auto sanitization via type inference [2], [28] come with limitations. Technological restrictions, such as the use of template languages, means such approaches are not widely applicable to many legacy web applications.

Johns et al. [29] have developed an abstract data type that strictly enforces data and code separation in a host language such as Java. However, this approach comes with a significant, 25%, run-time overhead. ScriptGard [28] is a run-time auto-sanitization technique in ASP.Net similar to Haldar et. al [30] for Java and WASP for SQL injection [20]. Advantages of these approaches are that they can automatically sanitize large scale legacy systems using a path-sensitive approach using binary code instrumentation. All these approaches require a runtime component that could incur runtime overheads. Furthermore, requiring a runtime component necessitates changes to existing infrastructures, such as browsers. Our approach works with all web languages (HTML, CSS, JavaScript) and requires no runtime support.

Static analysis techniques are widely used to detect XSS vulnerabilities using taint analysis techniques [31]–[35]. The main disadvantage of static analysis is high false positive rates [3]. Furthermore, static analysis tools can only check for the existence of the sanitization functions and not evaluate their effectiveness [36]. This limits the capability of static analysis to address the context-sensitiveness of sanitization errors [37].

Dynamic analysis techniques aim to evaluate application responses to detect any sanitization mistakes [38]–[41]. In the case of the black box testing, different algorithms such as combinatorial testing [42], [43], pattern-based algorithms [44], [45], and attack repositories [33] have been explored. Duchene et al. [15], [44] proposed a control- and data- flow aware fuzzing technique. They use a state-aware crawler to record application requests and responses and use them to infer an application's control and data flow. The fuzzing process is then guided by this information. The advantage of data flow inference is that it enables more accurate detection of stored XSS vulnerabilities. Because crawling-based inference is source-code independent and it uses automatic form filling and pruning techniques, the inferred control flow may not be complete, leading to potentially high false negatives. In contrast, we utilize source code analysis to extract all execution paths to avoid missing any sinks.

McAllister et. al [46] proposed an interactive black-box vulnerability scanner. They aim to increase test coverage by leveraging user activities through guided fuzzing. However, relying on user activities to increases test coverage is not complete and can lead to false negatives.

There are other types of XSS sanitization functions (other than encoders) not addressed by our approach. Consider a blogging web site allowing the use of HTML markup tags as input. HTML encoding functions are not proper here because they would disable all HTML markup tags. There are heuristic filters, e.g. [25], that try to block unwanted JavaScript programs in HTML body context. Such filters are difficult to verify automatically.

**Vulnerability Repair:** Various patch generation and vulnerability repair mechanisms have been developed by researchers. Medeiros et al. introduced a vulnerability detection and correction technique based on static analysis and machine learning. They define a fix as adding an encoder instead of replacing an an existing encoder, which can have side effects on the logic of the code such as over encoding [47].

Yu et al. [48] proposed an automata-based input validation mechanism for web applications. Given predefined attack signatures and benign input patterns (e.g, date, URLs, IP addresses) they check the input strings to find malicious patterns. Once an input matches a malicious signatures they generate encoding patches to modify the inputs . This approach does not necessarily prevent XSS attacks because encoding is context-sensitive and no context detection is provided in this work. Other approaches have addressed different vulnerabilities. For example, FixMeUp [49] is an access-control repair tool for web applications based on static-analysis. Ma et al. introduced CDRep [50] as an automatic repair tool for cryptographic defects in Android applications. They use decompilation and fault identification to detect cryptographic misuse defects.

Pattern-based Automatic program Repair(PAR) [51] is a general software repair mechanism that learns from previous code repairs (manually done by developers) to automatically generate code patches. It is similar to GenProg [52] by Weimer et al. which uses a genetic-programming-based approach for patch generation. Learning from examples is the technique used in VuRLE [53] as an automatic vulnerability detection and repair tool.

None of these vulnerability repair techniques are applicable to repair XSS vulnerabilities due to encoder misuse. Our approach goes further then repairing by verifying suggested repairs by unit testing.

## IX. CONCLUSION AND FUTURE WORK

In summary, we propose a unit testing approach to detect and repair cross-site scripting vulnerabilities caused by incorrect encoder usage. This approach can be easily integrated into existing software development practices and can pinpoint the location of a vulnerability in the source code. It can help developers find and fix XSS vulnerabilities early in the development cycle, when they unit test their code, without involving security experts. The grammar-based attack generation is a structured way to generate XSS attack strings. We were able to generate tests for vulnerabilities missed by popular attack repositories. More importantly, our grammar models can be modified to cover unknown or new attack scenarios. For example, a new version of a browser may offer new ways for attackers. Our approach also has low false positive rates. Our evaluation suggests that our unit-test based approach to detect XSS vulnerabilities is computationally feasible for large applications and can detect vulnerabilities that cannot be found using black-box fuzzing systems.

Our evaluation also suggest that the auto-fixing technique described in this paper can repair many XSS vulnerabilities in web applications by replacing erroneous encoders without making structural changes to the program.

This work can be extended in a number of ways. We plan to extend auto-repair to include code restructuring, addressing one of the limitations of our current approach. More evaluation with open source projects is also needed to further validate and improve our approach. We plan to extend our work to handle security sinks in client-side code that use asynchronous calls to web services and JSON-based communications. This extension requires JavaScript static analysis to find and repair vulnerabilities in hybrid mobile applications as well.

## Acknowledgment

This work is supported in part by the following grants from the National Science Foundation: 1129190, 1318854.

**Bei-Tseng Chu** Bei-Tseng "Bill" Chu received his Ph.D. in Computer Science from the University of Maryland at College Park. He is currently a Professor in the Department of Software and Information Systems at the University of North Carolina Charlotte. His research interests include software security, cyber threat intelligence, cybersecurity automation, and cybersecurity education.

**Heather Richter Lipford** Dr. Heather Richter Lipford is a Professor at the University of North Carolina at Charlotte. She completed her Ph.D. from the College of Computing at the Georgia Institute of Technology. Her research interests include usable privacy and security, human computer interaction, and security education.

**Mahmoud Mohammadi** Mahmoud Mohammadi is PhD student in the Department of Software and Information Systems at the University of North Carolina Charlotte. He received his master in information and communication technology from Tarbiat Modares university. His research interests are cognitive security, data privacy and adversarial machine learning.